\documentclass[aps,prb,twocolumn,balancelastpage]{revtex4-1} 
\usepackage{amsmath}    % need for subequations
\usepackage{graphicx}   % for figures
\usepackage{subfigure}
\usepackage{hyperref}
\usepackage{breakurl}
\newcommand{\unit}[1]{\ensuremath{\, \mathrm{#1}}}
\begin{document}
\title{Understanding Pound-Drever-Hall locking using voltage controlled
radio-frequency oscillators: An undergraduate experiment
}
\author{C.~E.~Liekhus-Schmaltz} \email{cliekhus@uwaterloo.ca}   
\author{J.~D.~D.~Martin}
\affiliation{Department of Physics and Astronomy, University of Waterloo, Waterloo, Ontario, N2L 3G1, Canada
}

\date{\today}

\begin{abstract}
We have developed a senior undergraduate experiment that illustrates frequency stabilization techniques using radio-frequency electronics. The primary objective is to frequency stabilize a voltage controlled oscillator to a cavity resonance at $800\unit{MHz}$ using the Pound-Drever-Hall method. This technique is commonly applied to stabilize lasers at optical frequencies. By using only radio-frequency equipment it is possible to systematically study aspects of the technique more thoroughly, inexpensively, and free from eye hazards. Students also learn about modular radio-frequency electronics and basic feedback control loops. By varying the temperature of the resonator, students can determine the thermal expansion coefficients of copper, aluminum, and super invar.
\end{abstract}

\maketitle

\section{Introduction}

The Pound-Drever-Hall technique is commonly used to frequency stabilize lasers to optical cavity resonances. It was originally developed by Pound\cite{pound:1946} for the frequency stabilization of microwave oscillators, and adapted to the optical domain by Drever and Hall and others.\cite{drever:1983} In brief, the source to be stabilized is frequency modulated. A diode detector (a photodiode in the optical domain) detects the reflection of the modulated source from a cavity. If the source is slightly detuned from a resonance, the diode detector signal will contain a component at the modulation frequency. When the source is on resonance, no component is observed at the modulation frequency. By mixing the diode signal with the modulation source, we can obtain a suitable error signal for feedback control of the oscillator frequency (zero on resonance, positive on one side and negative on the other). High bandwidth and a large capture range have made this technique popular for laser frequency stabilization in research laboratories. The technique is now rarely used for the stabilization of lower-frequency (microwave) oscillators, where a variety of alternative techniques exist.

Black\cite{black:2001} has written a pedagogical article on the basic theory of the Pound-Drever-Hall technique and an undergraduate experiment has been developed by Boyd {\it et al.}\cite{boyd:1996} to demonstrate laser frequency stabilization using the technique. A detailed guide to its implementation in a research context is available in Ref.~\onlinecite{fox:2001}.

The availability of inexpensive modular radio-frequency (RF) components has allowed us to develop a senior undergraduate experiment which is similar in spirit to the optical implementation of Pound-Drever-Hall, but which uses RF electronics rather than optical equipment. The three main pieces of equipment are a commercial voltage controlled oscillator, a resonating cavity, and an integrating control circuit.

The essence of the Pound-Drever-Hall technique is the phase change in the cavity reflection coefficient as the frequency passes through a resonance. With RF electronics it is straightforward to directly observe this phase shift using an unmodulated source. This observation is the basis of the interferometric cavity locking techniques sometimes applied in the microwave regime.\cite{Ivanov} By directly observing the phase shift (which is difficult in the optical domain due to the short wavelengths), the basis of the Pound-Drever-Hall technique is reinforced. Radio frequency electronics also provide a systematic way to vary the extent of source modulation and the cavity coupling.

The experiment begins by observing the relation between the input voltage and the output frequency, which is known as the tuning curve of the voltage controlled oscillator. The cavity resonance is then observed by scanning the frequency of the voltage controlled oscillator and measuring the power reflected from the cavity. Different coupling conditions can also be tested at this time. The real and imaginary parts of the reflection coefficient are investigated by mixing the reflected signal with a phase-shifted portion of the original signal to create a dispersion-like error signal which can be used to frequency stabilize the voltage controlled oscillator. The modulation properties of the voltage controlled oscillator are then investigated, and the relation between the modulation voltage, the tuning curve, and the strength of the frequency sidebands is confirmed. Once the modulation is understood quantitatively, the Pound-Drever-Hall technique is implemented, and plots of the error signal as a function of the detuning of the oscillator from the cavity resonance are obtained.

In the final step the voltage controlled oscillator is frequency stabilized using the Pound-Drever-Hall error signal. Locking can be verified by changing the temperature of the cavity and recording the stabilized frequency change using a frequency counter. The relation between the frequency and temperature can be used to determine the linear thermal expansion coefficient of copper. By changing the inner conductor it is also possible to measure the expansion coefficients of aluminum and super invar. In the following sections we explain these aspects of this experiment in more detail.

\section{Resonating Cavity}

\begin{figure}[tb]
	\centering
	\subfigure{(a)
	\includegraphics[width=2.5cm]{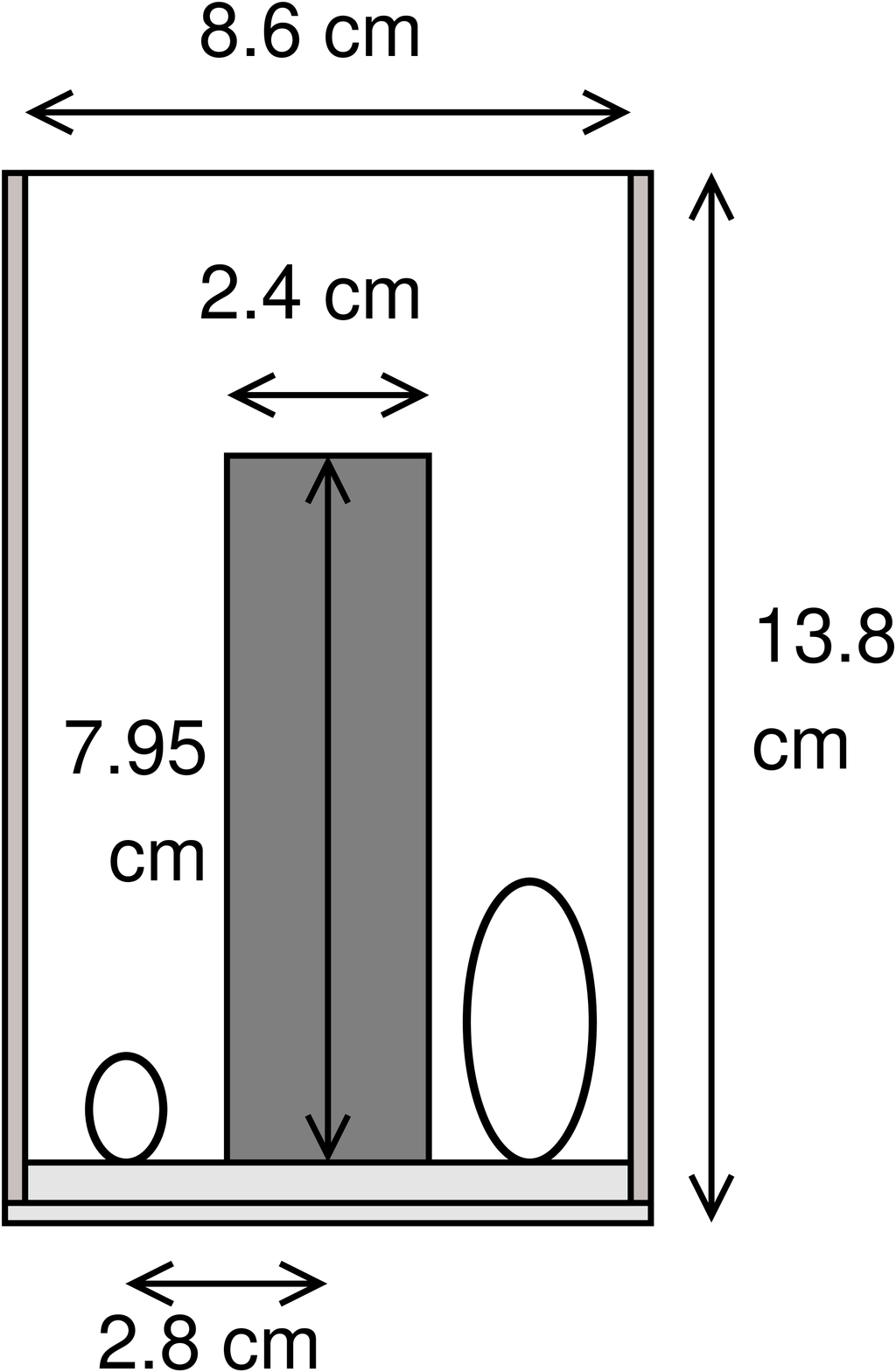}
	}
	\subfigure{(b)
	\includegraphics[width=3.5cm]{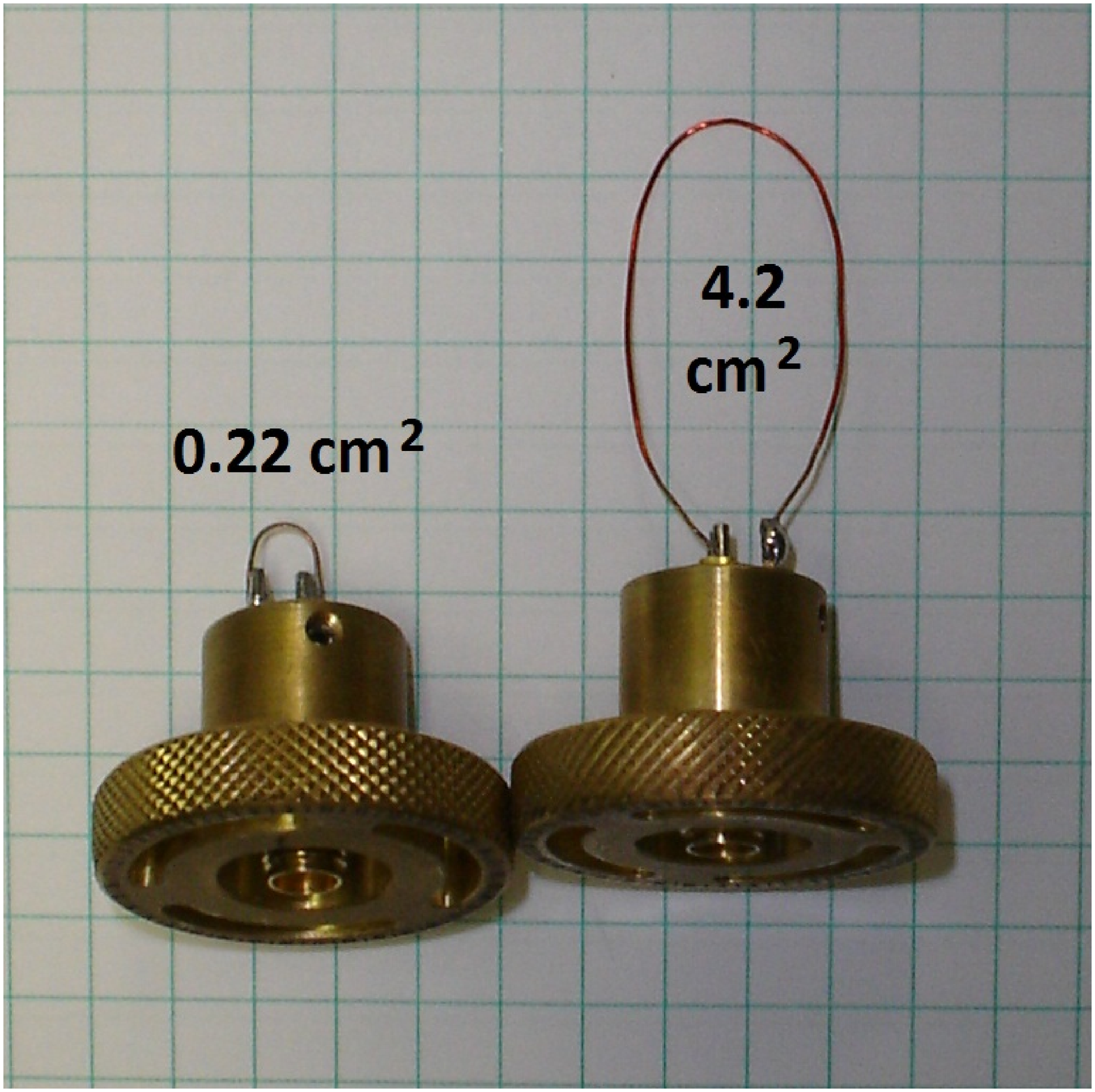}
	} \\
	\subfigure{(c)
	\includegraphics[width=3.5cm]{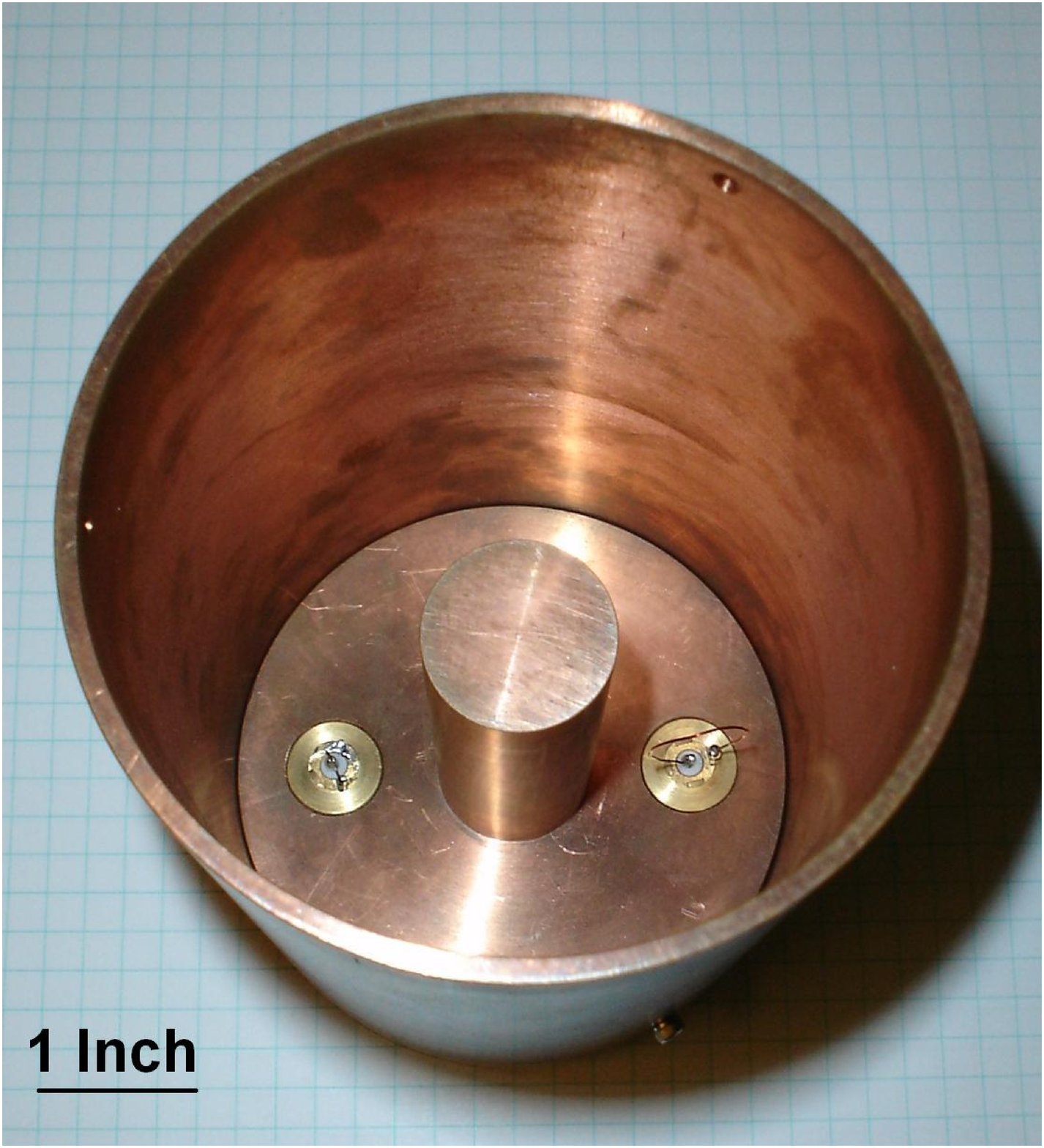}
	} 
	\subfigure{(d) 
	\includegraphics[width=3.5cm]{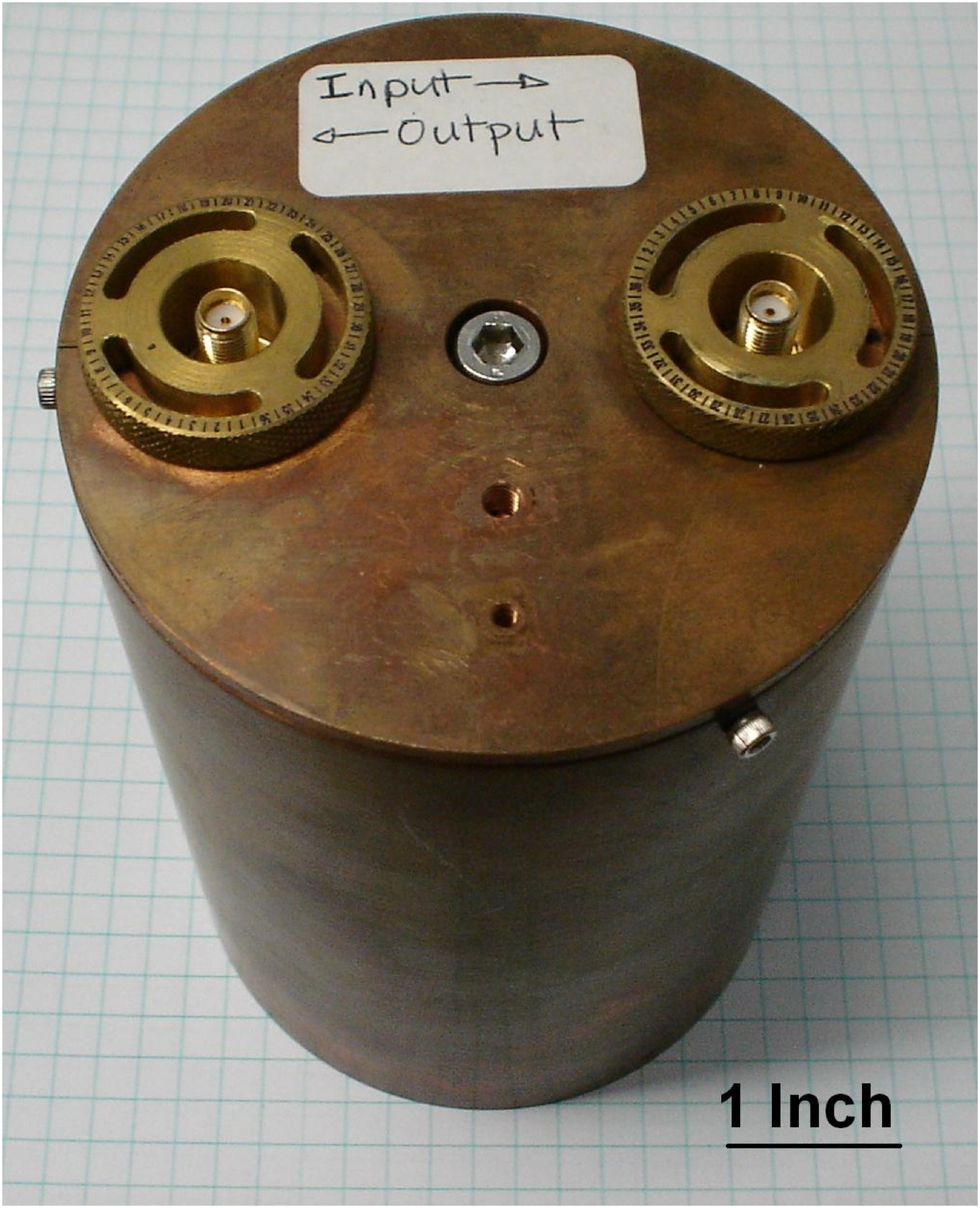}
	} 
\caption{$\lambda/4$ coaxial transmission line resonator. (a) Relevant dimensions. (b) The coupling loops inserted in the cartridges. (c) View from the open end (current node). (d) View of top of cavity (voltage node). The tops of the cartridges are seen here. }
\label{fig:Cavity}
\end{figure}

The resonating cavity shown in Fig.~\ref{fig:Cavity} consists of a $\lambda/4$ length coaxial transmission-line of $77\unit{\Omega}$ characteristic impedance, with one ended shorted and the other open-circuited.  Although we refer to it as a ``cavity,'' the current node end is left open allowing both visual inspection and the inner cylinder to be easily changed. The length of the inner cylinder is one quarter of the desired resonant wavelength corresponding to $800\unit{MHz}$. The resonant frequency is dictated by the availability of a suitable voltage controlled oscillator and cavity dimensions which are convenient for handling and inspection by students. The coaxial cavity type was chosen because it is similar to familiar resonating systems, such as transverse waves on strings. This configuration also allows us to observe the thermal expansion of the inner conductor because the resonant frequency is primarily determined by the length of the inner cylinder. The inner cylinder can also be changed to observe the thermal expansion of different materials.

Two holes for coupling loops are drilled in the cavity lid midway between the inner cylinder and the edge of the outer cylinder. The loops consist of 26 AWG copper wire attached to SMA connectors by soldering one end to the center pin and the other to ground. The SMA connectors are inserted into brass cartridges which fit into the holes in the lid (voltage node). The cartridges are labeled so that the angle of rotation can be read.

The two loops are approximately $0.3\unit{cm^2}$ and $4\unit{cm^2}$ in area. The larger loop is used to couple power into the cavity, and the smaller loop is used to detect power from the cavity. The size of the large loop is dictated by the requirement that under, critical, and over coupling be observable by rotating the loop cartridge. The other loop is small to reduce its impact on the quality factor.

The unloaded quality factor of this cavity, $Q_U \approx 700$, is small compared to literature values [Ref.~\onlinecite{terman1947radio}, Eq.~(70) gives $Q_U\approx 10800$]. We have constructed a similar cavity for research purposes\cite{cels2:2011} with a single threaded hole for an SMA-based coupling loop and have verified that the discrepancy is primarily due to the brass cartridges. However, the large, easily adjustable coupling loops, and the relatively small $Q_U$ are advantageous for this experiment.

\section{Resonance}

The experiment is based on a voltage controlled oscillator (Minicircuits, ZX95-850-S+) which has a central frequency which approximately matches the resonant frequency of the cavity ($\approx 800\unit{MHz}$)
and a modulation bandwidth much greater than the Pound-Drever-Hall modulation frequency ($\approx 10\unit{MHz}$). To observe the cavity resonance, a ramp function is applied to the tuning port of the voltage controlled oscillator while the output is connected to the large coupling loop through an isolator and then circulator. The signal that reflects from the cavity exits the circulator and is amplified before entering a detector diode followed by a 5\,k$\Omega$ resistor in parallel to ground. The diode voltage is observed using an oscilloscope.

The detector diode voltage to power relation was measured and is provided to the students. This relation depends on the load that the diode is driving. A 5\,k$\Omega$ parallel load resistor is used to ensure that the load is consistent between different oscilloscopes.

\begin{figure}[tb]
	\centering
		\includegraphics[width=8cm]{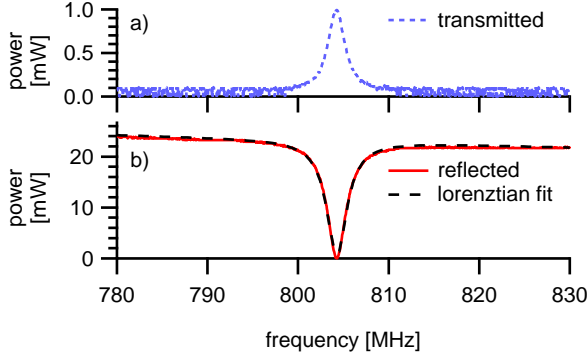}
\caption{Cavity resonance under critical coupling conditions. (a) Transmitted signal. (b) Reflected signal. The loaded quality factor $Q_L$ is determined by fitting a Lorentzian to the reflected power [see Eq.~\eqref{Gamma2}]. A linearly varying incident power has been included in the fit to accommodate for the frequency dependent losses of components other than the cavity.}
\label{fig:QDetermine}
\end{figure}

The students are asked to explore over, under, and critical coupling by varying the angle of the input coupling loops, with critical coupling being characterized as having the smallest reflected power on resonance. Once critical coupling is found and the loops are secured in this position, the reflected signal is analyzed to find the loaded quality factor, $Q_L$, of the cavity (see Fig.~\ref{fig:QDetermine}). To determine this value, a model for the reflection coefficient of the cavity must be determined.

The reflection coefficient of a one-port\cite{pozar2005microwave} is defined as $\Gamma_{\rm R} = \tilde v_-/\tilde v_+$, where $\tilde v_+$ and $\tilde v_-$ are phasors representing the incident and reflected traveling wave amplitudes at the location of the one-port. (We use $\tilde{~}$ to signify phasor quantities.) If an impedance $Z$ is driven through a transmission line of characteristic impedance $Z_0$, the reflection coefficient can be calculated to be
\begin{equation}
\Gamma_{\rm R}=\frac{Z-Z_0}{Z+Z_0}.
\label{eq:reflectcoeff}
\end{equation}

A cavity coupled to a transmission line can be modeled as a lumped element resonant circuit of total impedance $Z$ in the vicinity of a resonance, allowing its reflection coefficient to be calculated using Eq.~\eqref{eq:reflectcoeff}. Although the equivalence of the lumped circuit model can be established under quite general conditions,\cite{beringer:1948} a heuristic motivation specific to our situation will be given here.

\begin{figure}[tb]
	\centering
		\includegraphics[width=8cm]{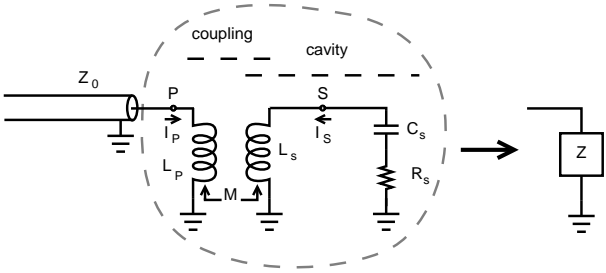}
                \caption{A lumped element circuit model of the input transmission line, coupling, and the resonating cavity.}
                \label{fig:Impedance}
\end{figure}

Near resonance, the voltage node end of a $\lambda/4$ coaxial resonator behaves like a series LCR resonant circuit -- a large amount of current flows for a small oscillating voltage applied between the inner and outer conductors. The input coupling loop interacts primarily with the oscillating magnetic field at this end, so we model the coupling using a non-ideal transformer, as shown in Fig.~\ref{fig:Impedance}. The secondary of the transformer is assumed to be part of the LCR resonator. (We ignore the second smaller loop in our cavity and assume that its contribution to cavity loss can be incorporated into $Q_U$.)

If we use the phasor relations
\begin{align}
\tilde{V}_P & = i \omega L_P \tilde{I}_P + i\omega M \tilde{I}_S \\
\tilde{V}_S &=i\omega L_S\tilde{I}_S+i\omega M \tilde{I}_P,
\label{inductor}
\end{align}
for the transformer (see Fig.~\ref{fig:Impedance}) where $M$ is the coupling coefficient, and
\begin{equation}
\tilde{V}_S=-\left(R_S + \frac{1}{i\omega C_S}\right) \tilde{I}_S,
\end{equation}
we find that $Z=\tilde{V}_P/\tilde{I}_P$ is given by
\begin{equation}
Z = i \omega L_P + \frac{\omega^2 M^2}{R_S+i\left(\omega L_S-\dfrac{1}{\omega C_S}\right)}.
\end{equation}
For a series LCR resonator $Q_U=(L_S/R_S)\omega_0$, where $\omega_0=1/\sqrt{L_SC_S}$ is the angular frequency at resonance.
We define $\Delta \omega=\omega-\omega_0$, assume $|\Delta \omega/\omega| \ll 1$, and simplify the expression for $Z$ as
\begin{equation}
Z \approx i\omega L_P + \left(\frac{\omega_0^2 M^2}{R_S}\right) \frac{1}{1+2 Q_U \dfrac{\Delta\omega}{\omega_0}i}.
\label{this}
\end{equation}
Equation~\eqref{this} can be interpreted as equivalent to the impedance of a parallel LCR circuit near resonance, with a resistance of $R_{\parallel}=\omega_0^2M^2/R_S$ in series with an inductor $L_{P}$. A coupling coefficient may be defined as\cite{ginzton1957microwave}
\begin{equation}
\kappa = \left( \dfrac{R_{\parallel}}{Z_0} \right) \dfrac{1}{1+\left(\omega L_P/Z_0\right)^2},
\end{equation}
so that by using Eqs.~\eqref{this} and \eqref{eq:reflectcoeff}, $\Gamma_{\rm R}$ may written as\cite{ginzton1957microwave}
\begin{equation}
\Gamma_{\rm R} = - \Gamma_{\rm L} \left(\dfrac{\kappa-1}{\kappa+1}\right) \dfrac{\left( 1-\dfrac{2 i \Delta \omega^{\prime} Q_U}{(\kappa-1)\omega_0}\right)}{\left( 1+\dfrac{2 i \Delta \omega^{\prime} Q_U}{(\kappa+1)\omega_0}\right)},
\label{Gamma}
\end{equation}
where $\Gamma_{\rm L}=(i\omega L_P-Z_0)/(i \omega L_P+Z_0)$ is a phase factor of unit magnitude, and $\Delta \omega^{\prime}=\Delta \omega - \omega \omega_0 \kappa L_P/(2Q_UZ_0)$. This frequency shift due to coupling is small, and we will assume that $\Delta \omega^{\prime} = \Delta \omega$.

We estimate the impedance of $L_P$ to have a magnitude of $300\unit{\Omega}$ at $800\unit{MHz}$,\cite{Grover} which is comparable to $Z_0$ ($50\unit{\Omega}$). Hence, $L_P$ contributes a significant phase to the overall reflection coefficient. This additional phase can be compensated for by introducing the appropriate phase change by an adjustable delay line. To simplify the following discussion of phase, we define a phase-shifted reflection coefficient $\Gamma=- \Gamma_{\rm R}/\Gamma_{\rm L}$. 

When looking at the reflected power we are interested in
\begin{equation}
|\Gamma_{\rm R}|^2=1-\dfrac{1-\left(\dfrac{\kappa-1}{\kappa+1}\right)^2}{1+4\left[\dfrac{Q_U}{(\kappa+1)}\dfrac{\Delta \omega}{\omega_0}\right]^2}.
\label{Gamma2}
\end{equation}
For critical coupling $\kappa=1$,
and $|\Gamma_{\rm R}|^2=0$ at resonance. Under coupling corresponds to $\kappa < 1$
and over coupling to $\kappa > 1$. If we define the loaded quality factor $Q_L \equiv \omega_0/\delta \omega$, where $\delta \omega$ is the full-width half-maximum of the resonance, we find from Eq.~\eqref{Gamma2} that $Q_L = Q_U/(\kappa+1)$. Thus critical coupling ($\kappa=1$) is a particularly convenient configuration for the determination of $Q_U$, and it is straightforward to experimentally identify ($|\Gamma_{\rm R}|^2=0$ at resonance; see Fig.~\ref{fig:QDetermine}). The rest of the experiment is done with critical coupling to simplify the derivations.

The expression for the reflection coefficient $\Gamma$ is analogous to the optical case,\cite{black:2001} provided that the optical cavity finesse is sufficiently high.

\section{Reflection coefficient}
\label{Sec:PhaseChange}

The Pound-Drever-Hall technique is sensitive to how the real and imaginary parts of the reflection coefficient $\Gamma$ vary with frequency near resonance. In particular, it is significant that the imaginary part of $\Gamma$ is anti-symmetric about the resonance, and falls to zero away from the resonance. In contrast, the real part of $\Gamma$ is symmetric about the resonance, and approaches $-1$ away from resonance.

\begin{figure}[tb]
	\centering
		\includegraphics[scale=.1]{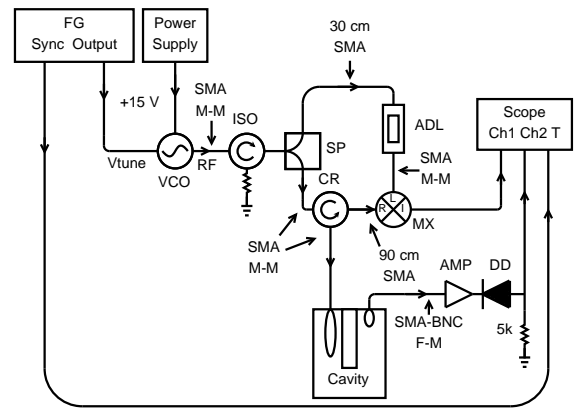}
\caption{Experimental setup to observe the real and imaginary parts of the cavity reflection coefficient. Depending on the bandwidth of the oscilloscope, it may be necessary to insert a low-pass filter after the mixer output. Key: ISO: isolator, SP: splitter, ADL: adjustable delay line, CR: circulator, MX: mixer, AMP: amplifier, DD: zero-bias Schottky diode, M-M: male to male connector, FG: function generator.}
\label{fig:part_c_setup}
\end{figure}

\begin{figure}[tb]
	\centering
		\includegraphics[width=8cm]{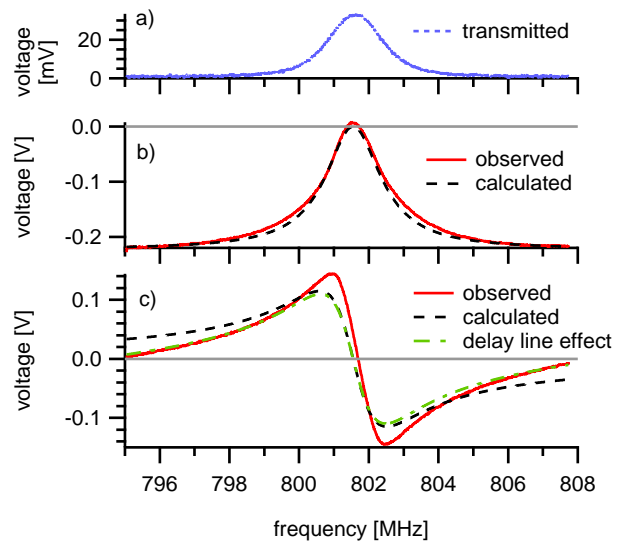}
\caption{Observation of cavity transmission and reflection using the setup of Fig.~\ref{fig:part_c_setup}. (a) Transmission through the cavity.
(b) Observation of the real part of the cavity reflection coefficient (to within a positive scale factor). (c) Observation of the imaginary part of the reflection coefficient (to within a positive scale factor). The curve labeled ``delay line effect'' is a calculation accounting for the variation in phase shift of the delay line with frequency. The calculations are vertically scaled for the best least squares fits.}
\label{fig:parts}
\end{figure}

Students can observe the imaginary and real parts of the reflection coefficient $\Gamma$ by mixing the reflected signal with a phase-shifted version of the incident signal (the reference). The technique is illustrated in Fig.~\ref{fig:part_c_setup}. An adjustable coaxial air delay line (General Radio Co.~874-LA) is used to set the relative phase between the reference and reflected signals. (The delay line can be replaced with a phase shifter if the instructions are modified to accommodate the phase shifter's mechanism for changing the phase.) With the loop detached from the cavity, the reference phase for the detection of $\text{Re}(\Gamma)$ can be set by adjusting the length to produce the largest negative dc output signal from the mixer. When the cavity is reattached, the mixer output will indicate $\text{Re}(\Gamma)$, as shown in Fig.~\ref{fig:parts}. When the length of the adjustable delay line is increased or decreased by $\lambda/4$, the mixer output will indicate $\text{Im}(\Gamma)$ or $-\text{Im}(\Gamma)$ respectively, which is also shown in Fig.~\ref{fig:parts}. The dispersion-like signal for $\text{Im}(\Gamma)$ is suitable as an error signal to frequency stabilize the voltage controlled oscillator (sometimes known as interferometric locking).\cite{Ivanov}

There are some discrepancies between the theoretical and observed reflection coefficients in Fig.~\ref{fig:parts}. The slight asymmetries are partially due to imperfect adjustment of the delay line. In addition, the asymptotic behavior of $\text{Im}(\Gamma)$ is influenced by the fact that the delay line is not a perfect, frequency-independent phase shifter. The phase shift variation with frequency can be calculated, and improves the agreement between the theory and observations, as shown in Fig.~\ref{fig:parts}. Although the theoretical reflection coefficient is calculated assuming that $\kappa=1$, we found that eliminating this assumption does not significantly improve agreement.

We note that due to the nature of the cavity design, the cartridges can rotate slightly or loosen while the setup is changed between measuring the resonance and the mixed signals, which causes the quality factor and/or the coupling constant to change. (We assume that the coupling constant $\kappa=1$.) In another design we tapped a hole directly into the lid of the cylinder so that there are no cartridges involved and the angles of the loops are fixed. This configuration might be more desirable, because it provides more reliable parameters for the theoretical calculation.

\section{Modulation of the voltage controlled oscillator}

Frequency modulation of the source to be stabilized (the voltage controlled oscillator in this case) is fundamental to the Pound-Drever-Hall technique. When a time-dependent voltage $V(t)=V_{\rm off}+V_{\rm amp} \cos ( \Omega t)$ is applied to the tuning port of the voltage controlled oscillator, we expect frequency modulation if $\Omega$ is within the voltage controlled oscillator's modulation bandwidth. We approximate the tuning curve of the voltage controlled oscillator by $f=f_0+A(V-V_{\rm off})$, and write the time-dependence of the frequency as $f(t)=f_0+\Delta f \cos (\Omega t)$, where $\Delta f = V_{\rm amp}A$. Because the phase is the time integral of the angular frequency $\phi = \!\int\! dt \, \omega(t)$, the output of the voltage controlled oscillator can be written in the phasor form:
\begin{equation}
\tilde{V}_{\rm inc}=\tilde{V}_0\,e^{i(\omega t+\beta \sin\Omega t)},
\label{mod}
\end{equation}
where $\beta = A V_{\rm amp}/[\Omega/(2\pi)]$ is the FM modulation index.

Equation \eqref{mod} can be rewritten using the Jacobi-Anger expansion
\cite{Arfken}
\begin{equation}
e^{i(\omega t +\beta \sin\Omega t)}= \sum_{m=-\infty}^{\infty} J_{m}(\beta)e^{i(\omega t+m\Omega t)},
\end{equation}
where $J_{m}(\beta)$ is the Bessel function of order $m$. For the assumption that $|\beta|\ll 1$,
\begin{equation}
\tilde{V}_{\rm inc} \approx \tilde{V}_0\big[J_0(\beta)e^{i\omega t}+J_1(\beta)e^{i(\omega+\Omega)t}-J_1(\beta)e^{i(\omega-\Omega)t}\big].
\label{Vinc}
\end{equation}
In this limit the output of the voltage controlled oscillator consists of three Fourier components with angular frequencies $\omega$, $\omega +\Omega$, and $\omega - \Omega$. The $\omega \pm \Omega$ components are called sidebands, and the central frequency is the carrier. The power in each component can be determined using the relation $P \propto |V|^2$. Therefore
$P_c=[J_0(\beta)]^2P_0$, and
$P_s=[J_1(\beta)]^2P_0$,
where $P_c$ is the power distributed to the carrier, $P_s$ is the power distributed to each sideband, and $P_0$ is the total power.

\begin{figure}[tb]
	\centering
		\includegraphics[width=8cm]{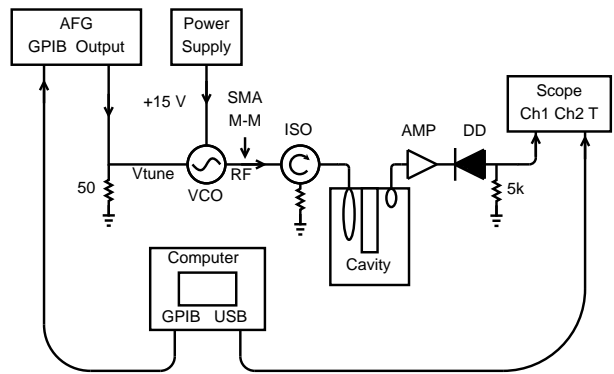}
\caption{Experimental setup for measuring Fourier components of the modulated voltage controlled oscillator. The relation between $\beta$ and the power in each Fourier component can be measured using this setup. The computer triggers the function generator to begin scanning and the oscilloscope to begin measuring data.
Key: ISO -  isolator, AMP - amplifier, DD - zero-bias Schottky diode, M-M - male to male connector.}
\label{fig:part_e_setup}
\end{figure}

It is desirable for students to confirm that modulation of the voltage controlled oscillator creates sidebands and that their powers have the expected dependence on $\beta$. This confirmation would usually be achieved using a relatively expensive RF spectrum analyzer. Alternatively, the cavity can be used as a transmission filter and the dc offset voltage of the voltage controlled oscillator scanned to observe the carrier and sideband powers (see Fig.~\ref{fig:part_e_setup}). Data collection is accomplished by computer control of a digital oscilloscope (Tektronix 2012C), and a function generator (Agilent 33120A). The function generator is used to apply both the modulating RF signal and dc offset $V_{\rm off}$ to the tuning port of the voltage controlled oscillator. A Python program (using the PyVisa module \cite{pyvisa:2011}) triggers the oscilloscope and then repetitively steps the dc offset of the function generator, thereby sweeping the carrier frequency. The output of the voltage controlled oscillator is sent to the cavity which acts as a filter. The signal transmitted through the cavity is amplified and then detected using a diode detector. As each of the Fourier components sweep through the resonant frequency, an increase in detector diode voltage is seen on the oscilloscope. The oscilloscope is setup to take only one trace per trigger so that students can read the detector diode voltages corresponding to the carrier and both sidebands directly off the oscilloscope display after the program ends. The observed diode voltages are then converted to power.

\begin{figure}[tb]
	\centering
		\includegraphics[width=8cm]{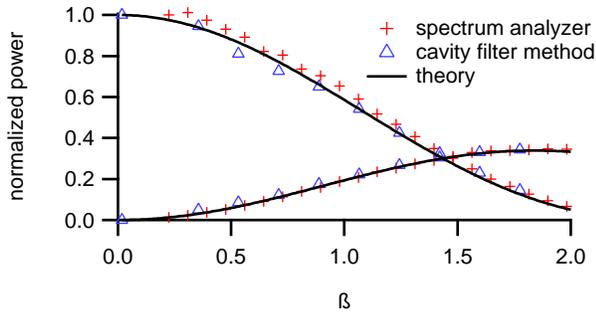}
\caption{Carrier and sideband powers as a function of the frequency modulation index $\beta$ observed using the cavity filter method, with $\Omega/(2\pi)= 6\unit{MHz}$, and an RF spectrum analyzer with $\Omega/(2\pi)= 10\unit{MHz}$. The values of $\beta$ are determined from the measured voltage controlled oscillator dc tuning curve. Also shown are the theoretically expected relations
$P_c=[J_0(\beta)]^2P_0$ and $P_s=[J_1(\beta)]^2P_0$ (see text).}
	\label{fig:SpectrumCavityCompare}
\end{figure}

This cavity filter method is suitable only if the powers of the carrier and sidebands remain unchanged as $V_{\rm off}$ is varied. As Fig.~\ref{fig:SpectrumCavityCompare} shows, both the cavity filter and RF spectrum analyzer methods show good agreement with the expected relation.

Larger values of $\beta$ show more features of the Bessel functions, and thus are more desirable. With the magnitude of $V_{\rm amp}$ constrained by the function generator output voltage and voltage controlled oscillator tuning port voltage limits, higher values of $\beta$ must be achieved by working with smaller $\Omega$. The lower bound on $\Omega$ is given by the bandwidth of the cavity filter. In this case we found that $\Omega/(2\pi) = 6 \unit{MHz}$ is a good compromise.

\section{Pound-Drever-Hall Locking}

Once the cavity reflection coefficient and frequency modulation of the voltage controlled oscillator have been studied, the students have a good basis for understanding the Pound-Drever-Hall technique. A derivation of the Pound-Drever-Hall signal has been given by Black,\cite{black:2001} which we briefly summarize here.
We begin by assuming that the voltage controlled oscillator has only three Fourier components (valid for small $\beta$). When the voltage controlled oscillator output reflects from the cavity, each frequency term will pick up a reflection coefficient $\Gamma$. Therefore, the total reflected power, $P_{\rm ref}\propto |V_{\rm ref}|^2$, is \cite{black:2001}
\begin{equation}
\begin{split}
P_{\rm ref}=& P_C|\Gamma(\omega)|^2+P_S{|\Gamma(\omega+\Omega)|^2 +P_S|\Gamma(\omega-\Omega)|^2} \\
& +2\sqrt{P_CP_S}\Big\{ \text{Re}[\Gamma(\omega)\Gamma^*(\omega+\Omega) \\
& -\Gamma^*(\omega)\Gamma(\omega-\Omega)]\cos ( \Omega t) +\text{Im}[\Gamma(\omega)\Gamma^*(\omega+\Omega)\\
& -\Gamma^*(\omega)\Gamma(\omega-\Omega)]\sin (\Omega t)\Big\}+2\Omega \, \text{terms}.
\end{split}
\label{eq:pdhdetec}
\end{equation}

It is useful to examine the situation when the voltage controlled oscillator frequency is close to the cavity resonance, that is, $\omega-\omega_0 \ll \Omega$. In most Pound-Drever-Hall implementations (including this experiment) it is normal that the modulation frequency is much greater than the cavity linewidth, that is,
$\Omega \gg \delta \omega$, so we can make the approximation (see Fig~\ref{fig:parts}), $\Gamma(\omega+\Omega)=\Gamma(\omega-\Omega)\approx -1$.
Therefore,
\begin{equation}
\text{Re}[\Gamma(\omega)\Gamma^*(\omega+\Omega) -\Gamma^*(\omega)\Gamma(\omega-\Omega)] \approx 0,
\end{equation}
and
\begin{equation}
\text{Im}[\Gamma(\omega)\Gamma^*(\omega+\Omega) -\Gamma^*(\omega)\Gamma(\omega-\Omega)] \approx -2i\text{Im}[\Gamma(\omega)].
\end{equation}
Both theoretically [Eq.~\eqref{Gamma}], and experimentally (Fig.~\ref{fig:parts}) we know that $\text{Im}[\Gamma(\omega)]$ is antisymmetric about the resonant frequency. Therefore it can be used as an error signal in a feedback loop to control the voltage controlled oscillator frequency. Its sign indicates whether the voltage controlled oscillator frequency should be lowered or raised to keep it matched with the cavity resonance.

In the diode output this desired error signal is modulated by $\sin (\Omega t)$, so it must be converted to dc and isolated from the rest of the terms in Eq.~\eqref{eq:pdhdetec}. This function can be performed by mixing the output of the diode with a $\sin(\Omega t)$ reference signal and subsequent filtering. The $\sin(\Omega t)$ reference can be obtained by splitting off a fraction of the voltage controlled oscillator modulation source output and applying an appropriate phase shift.

\begin{figure}[tb]
	\centering
		\includegraphics[width=8cm]{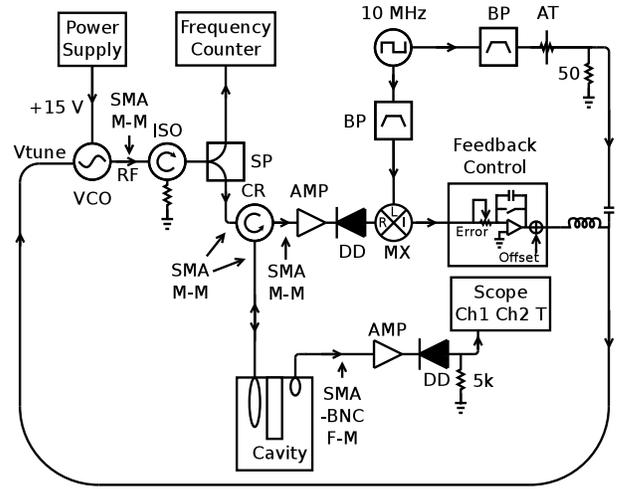}
\caption{The RF equivalent Pound-Drever-Hall locking method. The configuration shown is for locking the voltage controlled oscillator. The relative phases of the $10\unit{MHz}$ modulating and demodulating signals are set using coaxial cable lengths, which depend on the phase shifts of various components. To examine the Pound-Drever-Hall error signal as the voltage controlled oscillator frequency is tuned, as in Fig.~\ref{fig:PDHError}, the feedback control circuit is omitted. The voltage controlled oscillator offset is scanned by a function generator, and modulation applied through a bias T. The output of the mixer (I) is low-pass filtered and displayed on an oscilloscope. 
Key: ISO -  isolator, SP - splitter, CR - circulator, MX - mixer, AMP - amplifier, DD - zero-bias Schottky diode, M-M - male to male connector, BP - band pass filter for 10 MHz.}
\label{fig:part_g_setup}
\end{figure}

In optical implementations of Pound-Drever-Hall method, the reflected power from an optical cavity is detected by a photodiode, and Eq.~\eqref{eq:pdhdetec} is an expression for the photocurrent. In this all RF method we use a Schottky diode detector (Pasternack PE8000-50) for the same purpose (see Fig.~\ref{fig:part_g_setup}).

\begin{figure}[tb]
	\centering
		\includegraphics[width=8cm]{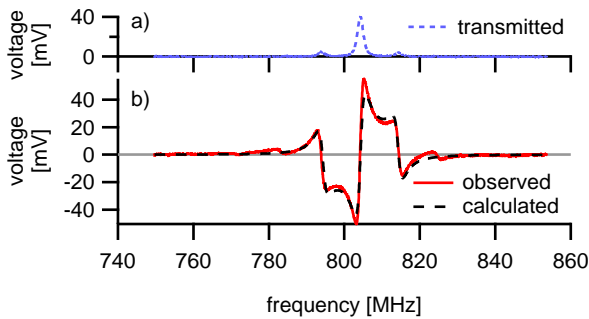}
\caption{Pound-Drever-Hall error signal as the voltage controlled oscillator frequency is tuned. (a) Transmission through the cavity. (b) Pound-Drever-Hall error signal for a voltage controlled oscillator modulation frequency of $\Omega/(2\pi)=10\unit{MHz}$ (see text for details).}
\label{fig:PDHError}
\end{figure}

To verify that the Pound-Drever-Hall method provides a suitable error signal students scan the voltage controlled oscillator frequency by applying a ramp to its tuning port, with RF modulation added through a bias T. A full scan, shown in Fig.~\ref{fig:PDHError}, shows the characteristic features of the Pound-Drever-Hall error signal.\cite{drever:1983} We can compare the observations to theory: $\text{Im}[\Gamma(\omega)\Gamma^*(\omega+\Omega) -\Gamma^*(\omega)\Gamma(\omega-\Omega)]$, where $\Gamma(\omega)$ is given by Eq.~\eqref{Gamma}. Using the previously determined resonance frequency, $Q_U$, and $\kappa=1$, the only fitting parameter required is an overall scale factor, provided $\beta$ is sufficiently small. As discussed by Black,\cite{black:2001} the optimal value of $\beta$ is 1.08. However, in the experiment a small $\beta$ ($\approx 0.3$) is used to reduce the magnitude of the higher-order features in the Pound-Drever-Hall spectrum.

There are four main factors which contribute to the difference between the theoretical model and the observed error signal. One is the phase of the $10\unit{MHz}$ reference, which if set incorrectly causes the slope on resonance to decrease. Second is a slight asymmetry of the sidebands (primarily caused by the frequency dependence of the circulator), which results in an asymmetry in the error signal. Third are second-order effects due to the second sidebands. This effect causes the zeros located $20\unit{MHz}$ from the resonant frequency. Lastly, the diode detector and mixer are not ideal devices.

We also note that errors in the determination of the quality factor and coupling factor, $\kappa$, might also be important, especially because changes in the setup can cause the coupling loops to shift slightly. Similar to the measurement of the real and imaginary parts of $\Gamma$, we have found that calculating the Pound-Drever-Hall error signal without assuming $\kappa=1$ does not improve agreement.

As noted, the interferometric $\text{Im}({\Gamma})$ signal provides a suitable error signal for frequency stabilization of the voltage controlled oscillator to the cavity. However, this method is not practical in the analogous optical case, where the relative signal and reference phase paths would have to be stabilized to within an optical wavelength. An advantage of the Pound-Drever-Hall technique is that the three Fourier components travel along the same path, and thus share a common phase. The phase difference between the reference and modulation at $\Omega$ is important, but significantly less demanding to control ($\Omega/(2\pi)=10\unit{MHz}$ here). In this context, it is useful to have students confirm --- by the use of long coaxial cables --- that the Pound-Drever-Hall signal depends on the relative phase of the $10\unit{MHz}$ signal, but that the sensitivity of the error signal to cable length is much less than for the interferometric setup (where an adjustable delay line was used to adjust the phase of the $800\unit{MHz}$ reference arm).

Once the error signal is observed, the students proceed to lock the voltage controlled oscillator using a simple integrator feedback control loop (see Fig.~\ref{fig:part_g_setup}) and can experiment with the influence of the gain and error signal polarity. While doing so, it is useful to monitor the power transmitted through the cavity to confirm locking.

An extension of this experiment could investigate the ability of the Pound-Drever-Hall technique to correct for rapid variations in the frequency of the source to be stabilized.\cite{black:2001} More specifically, phase noise suppression could be investigated by the injection of an additional noise modulation source into the tuning port of the voltage controlled oscillator.

\section{Measurement of Linear Thermal Expansion Coefficients}

It is possible to use the locked voltage controlled oscillator to find the linear thermal expansion coefficients of copper, aluminum, and super invar.  We begin by assuming that the approximate resonant frequency for a quarter wavelength coaxial cavity with an air dielectric is
\begin{equation}
f=\frac{4c}{L},
\label{eq:fforl}
\end{equation}
where $c$ is the speed of light and $L$ is the length of the cavity. The change in frequency as the temperature, $T$, changes is
\begin{equation}
\frac{df}{dT} = \frac{-4c}{L^2}\frac{dL}{dT} = \frac{-f}{L}\frac{dL}{dT}.
\label{eq:expand}
\end{equation}
Because $(dL/dT)/L$ is the linear expansion coefficient, Eq.~\eqref{eq:expand} expresses a relation between the resonant frequency and the linear expansion coefficient.

\begin{figure}[tb]
	\centering
		\includegraphics[width=3.375in]{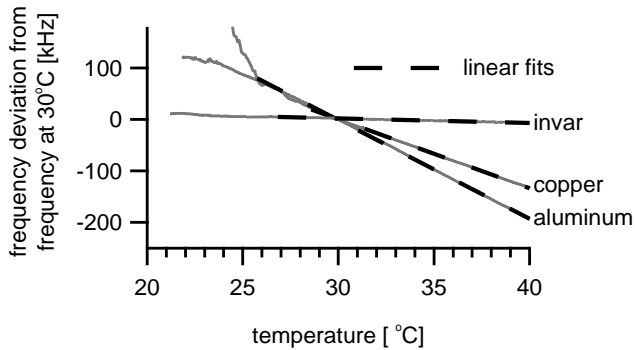}
\caption{Experimental data for determining the expansion coefficient of copper, aluminum, and super invar. By measuring the frequency change as the temperature of the cavity changes, we can determine the expansion of the metal which composes the inner cylinder. Also shown are the linear fits used for the determination of the linear coefficients of thermal expansion.}
\label{fig:Expansivity}
\end{figure}

When the voltage controlled oscillator is locked to the cavity, it will track the resonant frequency. Therefore, to find the linear expansion coefficient, we heat the cavity and measure the output frequency of the locked voltage controlled oscillator as well as the temperature while the system is in a Styrofoam box. The slope of a plot showing the frequency of the locked voltage controlled oscillator as the temperature changes and the average temperature provide the necessary information to determine the linear expansion coefficient according to Eq.~\eqref{eq:expand}. This procedure can be done for various materials by constructing the inner cylinder out of the desired material, provided its conductivity is sufficiently high. As shown in Fig.~\ref{fig:Expansivity} the thermal expansion coefficients of copper and aluminum are readily distinguished. It is also possible to observe the much lower thermal expansion coefficient of super invar (32-5 type,\cite{hightempmetals} with a silver plating of $5\unit{\mu m}$).

Figure \ref{fig:Expansivity} shows that the frequency-temperature relation is erratic at low temperatures when the heating is initially applied. This behavior is worse without an insulating Styrofoam box, and can be reduced by using a circulating fan. The erratic behavior is possibly related to initial temperature gradients and mechanical stresses. By heating the air rather than the cavity directly, we expect that these effects can be removed. This is not done in the undergraduate experiment due to time constraints.  If the air within the resonator is heated, we note that the influence on $c$ in Eq.~\eqref{eq:fforl} must be accounted for.\cite{smithweintraub}  The fractional change in frequency due to air temperature is on the order of $10^{-6}$ $^{\circ}$C$^{-1}$.

Once the frequency shows a well-defined relation to temperature, a linear fit is used to determine the relation between frequency and temperature which is used to calculate the coefficient of linear thermal expansion from Eq.~\eqref{eq:expand}. The results are shown in Table~\ref{tab:expand}.

\begin{table}[tb]
\centering
\begin{tabular}{lcc}
Material & Measured & Accepted \\
& ($10^{-6}$\,$^{\circ}$C$^{-1})$ & ($10^{-6}$\,$^{\circ}$C$^{-1})$\\ \hline
copper$^{\rm a}$ & 16.7 & 16.2 \\
aluminum$^{\rm a}$ & 23.9 & 23.4 \\
super invar$^{\rm b}$ & 1.12 & 0.63 \\
\hline
\end{tabular}
\caption{Values of linear thermal expansion coefficients. The accepted results for Cu and Al are from Ref.~\onlinecite{bechtell}; the accepted result for super invar is from Ref.~\onlinecite{hightempmetals}. \label{tab:expand}}
\end{table}

The measured linear thermal expansion coefficients are systematically larger than the accepted values by approximately $0.5 \times 10^{-6}$\,$^{\circ}$C$^{-1}$, possibly due to inadequate insulation of the cavity temperature sensing element. This effect could be reduced by heating the air surrounding the cavity, rather than the cavity directly.

\section{Implementation}

The total cost of implementing this experiment was approximately \$7000. All items were bought new, with the exception of the frequency counter and adjustable delay line (both of these were obtained from used test-equipment dealers). Many of the components employed are generic, and may be available in a standard undergraduate physics laboratory (frequency counter, oscilloscope, and function generator). Fabrication of the resonator was straightforward, and requires access to a lathe, milling machine, and a drill press. The outer cylinder was cut from a tube of the required size to minimize the required machining. The brass cartridges were manufactured using a computer numerical control (CNC) mill; it is possible to create these using a conventional milling machine and a lathe if a CNC mill is not available. The super invar inner cylinder was silver plated by a local shop for \$250.

\section{Concluding remarks}

To date, this experiment has been performed by six groups of undergraduates at the University of Waterloo. To complete the entire experiment typically takes two sessions of approximately four hours each. In an abbreviated single session the voltage controlled oscillator can be locked to the cavity using the interferometric technique, and thermal expansion measured using this lock. We omit an investigation of voltage controlled oscillator modulation and the Pound-Drever-Hall error signal.

The Pound-Drever-Hall technique is primarily confined to use in laser physics. A broader appeal of the experiment is that students gain familiarity with using modular RF components such as mixers and splitters. To assist the students with minimal direct involvement we have developed enhanced web-based apparatus diagrams,\cite{rfexpwebsite:2011} which students consult when doing the experiment. As a cursor is moved over the components in a diagram such as Fig.~\ref{fig:part_g_setup}, a photograph of the physical device appears, together with the manufacturer's part number and links to additional information.

Although designed for undergraduates, this experiment is also useful for new graduate students and researchers who are interested in learning about Pound-Drever-Hall locking and locking to optical cavities in general. For example, interferometric observation of the reflection phase shift (see Fig.~\ref{fig:parts}) provides insight into the H\"{a}nsch-Couillard locking technique.\cite{hansch:1980}

\begin{acknowledgements}
We gratefully acknowledge the assistance of Zhenwen Wang, J.\ Szubra, and H.\ Haile of the University of Waterloo Science Technical Services. We thank C.\ Bennett, J.\ Carter, S.\ De Young, and A.\ Lupascu for comments on the manuscript. This work was supported by the Natural Sciences and Engineering Research Council of Canada. 
\end{acknowledgements}

\pagebreak

\end{document}